%Paper: hep-th/9406072
%From: ramos@qmchep.DNET.NASA.GOV (Eduardo Ramos)
%Date: Mon, 13 Jun 94 15:17:44 -0400

%%%%%%%%%%%%%%%%%%%%%%%%%%%%%%%%%%%%%%%%%%%%%%%%%%%%%%%%%%%%%
%%%These are the macros for submission of papers to hep-th%%%
%%%The default setting is 12pt and 1 page/side but in the%%%%
%%%future it may allow people to choose also 10 pt and%%%%%%%
%%%2 pages/side.%%%%%%%%%%%%%%%%%%%%%%%%%%%%%%%%%%%%%%%%%%%%%
%%%%%%%%%%%%%%%%%%%%%%%%%%%%%%%%%%%%%%%%%%%%%%%%%%%%%%%%%%%%%
%
\def\unlockat{\catcode`\@=11}
\def\lockat{\catcode`\@=12}
\unlockat
\def\d@f@ult{} \newif\ifamsfonts \newif\ifafour
\def\m@ssage{\immediate\write16}  \m@ssage{}
\m@ssage{hep-th preprint macros.  Last modified 16/10/92 (jmf).}
\message{These macros work with AMS Fonts 2.1 (available via ftp from}
\message{e-math.ams.com).  If you have them simply hit "return"; if}
\message{you don't, type "n" now: }
\endlinechar=-1  %don't add spaces at end of line
\read-1 to\@nswer
\endlinechar=13
\ifx\@nswer\d@f@ult\amsfontstrue
    \m@ssage{(Will load AMS fonts.)}
\else\amsfontsfalse\m@ssage{(Won't load AMS fonts.)}\fi
\message{The default papersize is A4.  If you use US 8.5" x 11"}
\message{type an "a" now, else just hit "return": }
\endlinechar=-1  %don't add spaces at end of line
\read-1 to\@nswer
\endlinechar=13
\ifx\@nswer\d@f@ult\afourtrue
    \m@ssage{(Using A4 paper.)}
\else\afourfalse\m@ssage{(Using US 8.5" x 11".)}\fi
\nonstopmode
%
%%%%%%%%%%%%%%%%%%%%%%
%%%Font definitions%%%
%%%%%%%%%%%%%%%%%%%%%%
%

\font\twelverm=cmr12
\font\ninerm=cmr9
\font\sixrm=cmr6
\font\fourteenbf=cmbx12 scaled\magstep1
\font\twelvebf=cmbx12
\font\ninebf=cmbx9
\font\sixbf=cmbx6
\font\fourteeni=cmmi12 scaled\magstep1      \skewchar\fourteeni='177
\font\twelvei=cmmi12                        \skewchar\twelvei='177
\font\ninei=cmmi9                           \skewchar\ninei='177
\font\sixi=cmmi6                            \skewchar\sixi='177
\font\fourteensy=cmsy10 scaled\magstep2     \skewchar\fourteensy='60
\font\twelvesy=cmsy10 scaled\magstep1       \skewchar\twelvesy='60
\font\ninesy=cmsy9                          \skewchar\ninesy='60
\font\sixsy=cmsy6                           \skewchar\sixsy='60
\font\fourteenex=cmex10 scaled\magstep2
\font\twelveex=cmex10 scaled\magstep1

\ifamsfonts
   \font\ninex=cmex9
   
   \font\sixex=cmex7 at 6pt
   
\else
   \font\ninex=cmex10 at 9pt
   
   \font\sixex=cmex10 at 6pt
   
\fi
\font\fourteensl=cmsl10 scaled\magstep2
\font\twelvesl=cmsl10 scaled\magstep1

\font\sevensl=cmsl10 at 7pt
\font\sixsl=cmsl10 at 6pt

\font\fourteenit=cmti12 scaled\magstep1
\font\twelveit=cmti12

\font\fourteentt=cmtt12 scaled\magstep1
\font\twelvett=cmtt12
\font\fourteencp=cmcsc10 scaled\magstep2
\font\twelvecp=cmcsc10 scaled\magstep1

\ifamsfonts
   
\else
   
\fi
\newfam\cpfam
\font\fourteenss=cmss12 scaled\magstep1
\font\twelvess=cmss12
\font\tenss=cmss10
\font\niness=cmss9

\font\sevenss=cmss8 at 7pt
\font\sixss=cmss8 at 6pt
\newfam\ssfam
\newfam\msafam \newfam\msbfam \newfam\eufam
\ifamsfonts
 \font\fourteenmsa=msam10 scaled\magstep2
 \font\twelvemsa=msam10 scaled\magstep1
 \font\tenmsa=msam10
 \font\ninemsa=msam9
 \font\sevenmsa=msam7
 \font\sixmsa=msam6
 \font\fourteenmsb=msbm10 scaled\magstep2
 \font\twelvemsb=msbm10 scaled\magstep1
 \font\tenmsb=msbm10
 \font\ninemsb=msbm9
 \font\sevenmsb=msbm7
 \font\sixmsb=msbm6
 \font\fourteeneu=eufm10 scaled\magstep2
 \font\twelveeu=eufm10 scaled\magstep1
 \font\teneu=eufm10
 \font\nineeu=eufm9
 
 \font\seveneu=eufm7
 \font\sixeu=eufm6
 \def\hexnumber@#1{\ifnum#1<10 \number#1\else
  \ifnum#1=10 A\else\ifnum#1=11 B\else\ifnum#1=12 C\else
  \ifnum#1=13 D\else\ifnum#1=14 E\else\ifnum#1=15 F\fi\fi\fi\fi\fi\fi\fi}
 \def\hexmsa{\hexnumber@\msafam}
 \def\hexmsb{\hexnumber@\msbfam} 
\fi
\newdimen\b@gheight             \b@gheight=12pt
\newcount\f@ntkey               \f@ntkey=0
\def\f@m{\afterassignment\samef@nt\f@ntkey=}
\def\samef@nt{\fam=\f@ntkey \the\textfont\f@ntkey\relax}
\def\rm{\f@m0 }
\def\mit{\f@m1 }
\def\cal{\f@m2 }
\def\it{\f@m\itfam}
\def\sl{\f@m\slfam}
\def\bf{\f@m\bffam}
\def\tt{\f@m\ttfam}
\def\caps{\f@m\cpfam}
\def\ssf{\f@m\ssfam}
\ifamsfonts
 \def\msa{\f@m\msafam}
 \def\msb{\f@m\msbfam} \let\bb=\msb
 \def\eu{\f@m\eufam}
\else
 \let \bb=\bf \let\eu=\bf
\fi
\def\fourteenpoint{\relax
    \textfont0=\fourteencp          \scriptfont0=\tenrm
      \scriptscriptfont0=\sevenrm
    \textfont1=\fourteeni           \scriptfont1=\teni
      \scriptscriptfont1=\seveni
    \textfont2=\fourteensy          \scriptfont2=\tensy
      \scriptscriptfont2=\sevensy
    \textfont3=\fourteenex          \scriptfont3=\twelveex
      \scriptscriptfont3=\tenex
    \textfont\itfam=\fourteenit     \scriptfont\itfam=\tenit
    \textfont\slfam=\fourteensl     \scriptfont\slfam=\tensl
      \scriptscriptfont\slfam=\sevensl
    \textfont\bffam=\fourteenbf     \scriptfont\bffam=\tenbf
      \scriptscriptfont\bffam=\sevenbf
    \textfont\ttfam=\fourteentt
    \textfont\cpfam=\fourteencp
    \textfont\ssfam=\fourteenss     \scriptfont\ssfam=\tenss
      \scriptscriptfont\ssfam=\sevenss
    \ifamsfonts
       \textfont\msafam=\fourteenmsa     \scriptfont\msafam=\tenmsa
         \scriptscriptfont\msafam=\sevenmsa
       \textfont\msbfam=\fourteenmsb     \scriptfont\msbfam=\tenmsb
         \scriptscriptfont\msbfam=\sevenmsb
       \textfont\eufam=\fourteeneu     \scriptfont\eufam=\teneu
         \scriptscriptfont\eufam=\seveneu \fi
    \samef@nt
    \b@gheight=14pt
    \setbox\strutbox=\hbox{\vrule height 0.85\b@gheight
                                depth 0.35\b@gheight width\z@ }}
\def\twelvepoint{\relax
    \textfont0=\twelverm          \scriptfont0=\ninerm
      \scriptscriptfont0=\sixrm
    \textfont1=\twelvei           \scriptfont1=\ninei
      \scriptscriptfont1=\sixi
    \textfont2=\twelvesy           \scriptfont2=\ninesy
      \scriptscriptfont2=\sixsy
    \textfont3=\twelveex          \scriptfont3=\ninex
      \scriptscriptfont3=\sixex
    \textfont\itfam=\twelveit    %\scriptfont\itfam=\nineit
    \textfont\slfam=\twelvesl    %\scriptfont\slfam=\ninesl
      \scriptscriptfont\slfam=\sixsl
    \textfont\bffam=\twelvebf     \scriptfont\bffam=\ninebf
      \scriptscriptfont\bffam=\sixbf
    \textfont\ttfam=\twelvett
    \textfont\cpfam=\twelvecp
    \textfont\ssfam=\twelvess     \scriptfont\ssfam=\niness
      \scriptscriptfont\ssfam=\sixss
    \ifamsfonts
       \textfont\msafam=\twelvemsa     \scriptfont\msafam=\ninemsa
         \scriptscriptfont\msafam=\sixmsa
       \textfont\msbfam=\twelvemsb     \scriptfont\msbfam=\ninemsb
         \scriptscriptfont\msbfam=\sixmsb
       \textfont\eufam=\twelveeu     \scriptfont\eufam=\nineeu
         \scriptscriptfont\eufam=\sixeu \fi
    \samef@nt
    \b@gheight=12pt
    \setbox\strutbox=\hbox{\vrule height 0.85\b@gheight
                                depth 0.35\b@gheight width\z@ }}
\twelvepoint
%
%%%%%%%%%%%%%%%%%
%%%Basic skips%%%
%%%%%%%%%%%%%%%%%
%
\baselineskip = 15pt plus 0.2pt minus 0.1pt %was 20pt ...
\lineskip = 1.5pt plus 0.1pt minus 0.1pt
\lineskiplimit = 1.5pt
\parskip = 6pt plus 2pt minus 1pt
\interlinepenalty=50
\interfootnotelinepenalty=5000
\predisplaypenalty=9000
\postdisplaypenalty=500
\hfuzz=1pt
\vfuzz=0.2pt
\dimen\footins=24 truecm % 8 truein in SB
\ifafour
 \hsize=16cm \vsize=22cm
\else
 \hsize=6.5in \vsize=9in
\fi
%
%%%%%%%%%%%%%%%
%%%Footnotes%%%
%%%%%%%%%%%%%%%
%
\skip\footins=\medskipamount
\newcount\fnotenumber
\def\clearfnotenumber{\fnotenumber=0} \clearfnotenumber
\def\fnote{\global\advance\fnotenumber by1 \generatefootsymbol
 \footnote{$^{\footsymbol}$}}
\def\fd@f#1 {\xdef\footsymbol{\mathchar"#1 }}
\def\generatefootsymbol{\iffrontpage\ifcase\fnotenumber
\or \fd@f 279 \or \fd@f 27A \or \fd@f 278 \or \fd@f 27B
\else  \fd@f 13F \fi
\else\xdef\footsymbol{\the\fnotenumber}\fi}
%
%%%%%%%%%%%%%%%%%%%%%%%%%%%%%
%%%Sections and Appendices%%%
%%%%%%%%%%%%%%%%%%%%%%%%%%%%%
%
\newcount\secnumber \newcount\appnumber
\def\clearappnumber{\appnumber=64} \def\clearsecnumber{\secnumber=0}
\clearsecnumber \clearappnumber
\newif\ifs@c % this is true if within a section as opposed to an appendix
\newif\ifs@cd % this is true if the article is being section'd
\s@cdtrue % this is the default
\def\unsectioned{\s@cdfalse\let\section=\subsection}
\newskip\sectionskip         \sectionskip=\medskipamount
\newskip\headskip            \headskip=8pt plus 3pt minus 3pt
\newdimen\sectionminspace    \sectionminspace=10pc
\def\Titlestyle#1{\par\begingroup \interlinepenalty=9999
     \leftskip=0.02\hsize plus 0.23\hsize minus 0.02\hsize
     \rightskip=\leftskip \parfillskip=0pt
     \advance\baselineskip by 0.5\baselineskip%this is a test...
     \hyphenpenalty=9000 \exhyphenpenalty=9000
     \tolerance=9999 \pretolerance=9000
     \spaceskip=0.333em \xspaceskip=0.5em
     \fourteenpoint
  \noindent #1\par\endgroup }
\def\titlestyle#1{\par\begingroup \interlinepenalty=9999
     \leftskip=0.02\hsize plus 0.23\hsize minus 0.02\hsize
     \rightskip=\leftskip \parfillskip=0pt
     \hyphenpenalty=9000 \exhyphenpenalty=9000
     \tolerance=9999 \pretolerance=9000
     \spaceskip=0.333em \xspaceskip=0.5em
     \fourteenpoint
   \noindent #1\par\endgroup }
\def\spacecheck#1{\dimen@=\pagegoal\advance\dimen@ by -\pagetotal
   \ifdim\dimen@<#1 \ifdim\dimen@>0pt \vfil\break \fi\fi}
\def\section#1{\cleareqnumber \s@ctrue \global\advance\secnumber by1
   \par \ifnum\the\lastpenalty=30000\else
   \penalty-200\vskip\sectionskip \spacecheck\sectionminspace\fi
   \noindent {\caps\enspace\S\the\secnumber\quad #1}\par
   \nobreak\vskip\headskip \penalty 30000 }
\def\undertext#1{\vtop{\hbox{#1}\kern 1pt \hrule}}
\def\subsection#1{\par
   \ifnum\the\lastpenalty=30000\else \penalty-100\smallskip
   \spacecheck\sectionminspace\fi
   \noindent\undertext{#1}\enspace \vadjust{\penalty5000}}

\def\appendix#1{\cleareqnumber \s@cfalse \global\advance\appnumber by1
   \par \ifnum\the\lastpenalty=30000\else
   \penalty-200\vskip\sectionskip \spacecheck\sectionminspace\fi
   \noindent {\caps\enspace Appendix \char\the\appnumber\quad #1}\par
   \nobreak\vskip\headskip \penalty 30000 }
\def\ack{\par\penalty-100\medskip \spacecheck\sectionminspace
   \line{\fourteencp\hfil ACKNOWLEDGEMENTS\hfil}%
\nobreak\vskip\headskip }
\def\refs{\begingroup \par\penalty-100\medskip \spacecheck\sectionminspace
   \line{\fourteencp\hfil REFERENCES\hfil}%
\nobreak\vskip\headskip \frenchspacing }
\def\endrefs{\par\endgroup}
%--- Note added
%
%%%%%%%%%%%%%%%%%%%%%%%%%%%%%%%%%
%%%Running heads and footlines%%%
%%%%%%%%%%%%%%%%%%%%%%%%%%%%%%%%%
%
\newif\iffrontpage \frontpagefalse
\headline={\hfil}
\footline={\iffrontpage\hfil\else \hss\twelverm
-- \folio\ --\hss \fi }
%
%%%%%%%%%%%%%%%%
%%%Title page%%%
%%%%%%%%%%%%%%%%
%
\newskip\frontpageskip \frontpageskip=12pt plus .5fil minus 2pt
\def\titlepage{\global\frontpagetrue\hrule height\z@ \relax
               \pubblock\relax }
\def\endtitlepage{\vfil\break\clearfnotenumber\frontpagefalse}
\def\title#1{\vskip\frontpageskip\Titlestyle{\caps #1}\vskip3\headskip}
\def\author#1{\vskip.5\frontpageskip\titlestyle{\caps #1}\nobreak}
\def\and{\par\kern 5pt \centerline{\sl and}}
\def\andauthor{\vskip.5\frontpageskip\centerline{and}\author}

\def\address#1{\par\kern 5pt\titlestyle{\it #1}}
\def\andaddress{\par\kern 5pt \centerline{\sl and} \address}

\def\abstract#1{\par\dimen@=\prevdepth \hrule height\z@ \prevdepth=\dimen@
   \vskip\frontpageskip\spacecheck\sectionminspace
   \centerline{\fourteencp ABSTRACT}\vskip\headskip
   {\noindent #1}}

\def\email#1{\fnote{\tentt e-mail: #1\hfill}}

%
%%%%%%%%%%%%%%%%%%%%
%%%some addresses%%%
%%%%%%%%%%%%%%%%%%%%
%

%

%
\def\QMW{\address{%
   Department of Physics, Queen Mary and Westfield College\break
   Mile End Road, London E1 4NS, UK}}
%

%
%%%%%%%%%%%%%%%%
%%%References%%%
%%%%%%%%%%%%%%%%
%
\newcount\refnumber \def\clearrefnumber{\refnumber=0}  \clearrefnumber
\newwrite\R@fs                              %This opens a file .refs with
\immediate\openout\R@fs=\jobname.refs %the references in order of
                                            %appearance.
\def\closerefs{\immediate\closeout\R@fs} %close file so that TeX can read it
\def\refsout{\closerefs\refs
\unlockat
\input\jobname.refs
\lockat
\endrefs}
\def\refitem#1{\item{{\bf #1}}}%just bolds it so that \bf does not expand
\def\ifundefined#1{\expandafter\ifx\csname#1\endcsname\relax}
\def\[#1]{\ifundefined{#1R@FNO}%
\global\advance\refnumber by1%
\expandafter\xdef\csname#1R@FNO\endcsname{[\the\refnumber]}%
\immediate\write\R@fs{\noexpand\refitem{\csname#1R@FNO\endcsname}%
\noexpand\csname#1R@F\endcsname}\fi{\bf \csname#1R@FNO\endcsname}}
\def\refdef[#1]#2{\expandafter\gdef\csname#1R@F\endcsname{{#2}}}
%
%%%%%%%%%%%%%%%
%%%Equations%%%
%%%%%%%%%%%%%%%
%
\newcount\eqnumber \def\cleareqnumber{\eqnumber=0}
\newif\ifal@gn \al@gnfalse  % this is true if within an \eqalignno
\def\veqnalign#1{\al@gntrue \vbox{\eqalignno{#1}} \al@gnfalse}
\def\eqnalign#1{\al@gntrue \eqalignno{#1} \al@gnfalse}
\def\(#1){\relax%
\ifundefined{#1@Q}
 \global\advance\eqnumber by1
 \ifs@cd
  \ifs@c
   \expandafter\xdef\csname#1@Q\endcsname{{%
\noexpand\rm(\the\secnumber .\the\eqnumber)}}
  \else
   \expandafter\xdef\csname#1@Q\endcsname{{%
\noexpand\rm(\char\the\appnumber .\the\eqnumber)}}
  \fi
 \else
  \expandafter\xdef\csname#1@Q\endcsname{{\noexpand\rm(\the\eqnumber)}}
 \fi
 \ifal@gn
    & \csname#1@Q\endcsname
 \else
    \eqno \csname#1@Q\endcsname
 \fi
\else%
\csname#1@Q\endcsname\fi\global\let\@Q=\relax}
%
%%%%%%%%%%%%%%%%%
%%%Mathematica%%%
%%%%%%%%%%%%%%%%%
%
\newif\ifm@thstyle \m@thstylefalse
\def\mathstyle{\m@thstyletrue}
\def\proclaim#1#2\par{\smallbreak\begingroup%        small --> med???
\advance\baselineskip by -0.25\baselineskip%
\advance\belowdisplayskip by -0.35\belowdisplayskip%
\advance\abovedisplayskip by -0.35\abovedisplayskip%
    \noindent{\caps#1.\enspace}{#2}\par\endgroup%
\smallbreak}%--- defs, thms, ...                     small --> med???
\def\m@kem@th<#1>#2#3{%
\ifm@thstyle \global\advance\eqnumber by1
 \ifs@cd
  \ifs@c
   \expandafter\xdef\csname#1\endcsname{{%
\noexpand #2\ \the\secnumber .\the\eqnumber}}
  \else
   \expandafter\xdef\csname#1\endcsname{{%
\noexpand #2\ \char\the\appnumber .\the\eqnumber}}
  \fi
 \else
  \expandafter\xdef\csname#1\endcsname{{\noexpand #2\ \the\eqnumber}}
 \fi
 \proclaim{\csname#1\endcsname}{#3}
\else
 \proclaim{#2}{#3}
\fi}
\def\Thm<#1>#2{\m@kem@th<#1M@TH>{Theorem}{\sl#2}}%--- Theorem
\def\Prop<#1>#2{\m@kem@th<#1M@TH>{Proposition}{\sl#2}}%--- Proposition
\def\Def<#1>#2{\m@kem@th<#1M@TH>{Definition}{\rm#2}}%--- Definition
\def\Lem<#1>#2{\m@kem@th<#1M@TH>{Lemma}{\sl#2}}%--- Lemma
\def\Cor<#1>#2{\m@kem@th<#1M@TH>{Corollary}{\sl#2}}%--- Corollary
\def\Conj<#1>#2{\m@kem@th<#1M@TH>{Conjecture}{\sl#2}}%--- Conjecture
\def\Rmk<#1>#2{\m@kem@th<#1M@TH>{Remark}{\rm#2}}%--- Remark
\def\Exm<#1>#2{\m@kem@th<#1M@TH>{Example}{\rm#2}}%--- Example
\def\Qry<#1>#2{\m@kem@th<#1M@TH>{Query}{\it#2}}%--- Query
%
%--- Proof
%

%
\def\<#1>{\csname#1M@TH\endcsname}
%
%%%%%%%%%%%%%%%%%%%
%%%Abbreviations%%%
%%%%%%%%%%%%%%%%%%%
%
\def\ref#1{{\bf [#1]}}%--- [ref]
%--- et al.
\def\ie{{\it i.e.\/}}%--- i.e.
%--- e.g.
%--- Cf.
%--- cf.
 %--- double left quote
%--- th as in fifth
\def\nl{\hfil\break}%--- new line
%
%%%%%%%%%%%%%%%%%
%%%Mathematics%%%
%%%%%%%%%%%%%%%%%
%
%--- def over =
%--- Halmos Q.E.D.

%--- implies
%--- is implied by
%--- if and only if
\def\lapprox{\hbox{\lower3pt\hbox{$\buildrel<\over\sim$}}}% approx lt
\def\gapprox{\hbox{\lower3pt\hbox{$\buildrel<\over\sim$}}}% approx gt
\def\quotient#1#2{#1/\lower0pt\hbox{${#2}$}}%--- factor objects
\def\fr#1/#2{\mathord{\hbox{${#1}\over{#2}$}}}
\ifamsfonts
 \mathchardef\empty="0\hexmsb3F %--- better empty set than \emptyset
 \mathchardef\lsemidir="2\hexmsb6E % semidirect |x
 \mathchardef\rsemidir="2\hexmsb6F % semidirect x|
\else
 \let\empty=\emptyset
 \def\lsemidir{\mathbin{\hbox{\hskip2pt\vrule height 5.7pt depth -.3pt
    width .25pt\hskip-2pt$\times$}}}
 \def\rsemidir{\mathbin{\hbox{$\times$\hskip-2pt\vrule height 5.7pt
    depth -.3pt width .25pt\hskip2pt}}}
\fi
%
%--- injective map
%--- surjective map
%--- bijective map
\def\to{\rightarrow}%--- mapping
%--- long mapping
%--- isom over -->
\def\lra{\leftrightarrow}%--- just an abbrev.
%

%
 %--- commutative diagram macro
 %--- map in complex
%
\def\reals{\mathord{\bb R}} %--- reals
 %--- complex nos.
 %--- quaternions
 %--- integers
 %--- rationals
 %--- naturals
 %--- ground field
%
%--- Hom(omorphisms)
%--- tr(ace)
%--- Tr(ace)
%--- End(omorphisms)
%--- Mor(phisms)
%--- Aut(omorphisms)
%--- aut(omorphisms)
%--- supertrace
%--- superdeterminant
%--- kernel
%--- cokernel
%--- image
%
\def\underrightarrow#1{\vtop{\ialign{##\crcr
      $\hfil\displaystyle{#1}\hfil$\crcr
      \noalign{\kern-\p@\nointerlineskip}
      \rightarrowfill\crcr}}} %--- modification of \overrightarrow
\def\underleftarrow#1{\vtop{\ialign{##\crcr
      $\hfil\displaystyle{#1}\hfil$\crcr
      \noalign{\kern-\p@\nointerlineskip}
      \leftarrowfill\crcr}}}  %--- modification of \overleftarrow

\def\comm#1#2{\left[#1\, ,\,#2\right]}%--- [ , ]
%--- { , }
%--- [ , }
%
%--- Lie derivative
%--- vartnl derivative
%--- partial derivative
%--- full derivative
%
%%%%%%%%%%%%%%
%%%Journals%%%
%%%%%%%%%%%%%%
%
\def\PRL#1#2#3{{\sl Phys. Rev. Lett.} {\bf#1} (#2) #3}
\def\NPB#1#2#3{{\sl Nucl. Phys.} {\bf B#1} (#2) #3}

\def\CMP#1#2#3{{\sl Comm. Math. Phys.} {\bf #1} (#2) #3}
\def\PRD#1#2#3{{\sl Phys. Rev.} {\bf D#1} (#2) #3}

\def\PLB#1#2#3{{\sl Phys. Lett.} {\bf #1B} (#2) #3}

\def\FAaIA#1#2#3{{\sl Functional Analysis and Its Application} {\bf #1} (#2)
#3}

\def\JPA#1#2#3{{\sl J. Physics} {\bf A#1} (#2) #3}

\def\MPLA#1#2#3{{\sl Mod. Phys. Lett.} {\bf A#1} (#2) #3}

\def\JETPL#1#2#3{{\sl  Sov. Phys. JETP Lett.} {\bf #1} (#2) #3}

\lockat

\def\W{{\ssf W}}
\def\d{\partial}
\def\L{{\ssf L}}
\def\KdV{{\ssf KdV}}
\def\dddot#1{\hbox{$\mathop{#1}\limits^{\ldots}$}}
\def\ddxperp{\ddot x_\perp}

\def\dqperp{\dot q_\perp}

\unsectioned
%
%
%%%%%%%%%%%%%%%%%%%%%%%%%%%%
% These are the references
%%%%%%%%%%%%%%%%%%%%%%%%%%%%
\refdef[extgeo]{G. Sotkov and M. Stanishkov, \NPB{356}{1991}{439};
G. Sotkov, M. Stanishkov and C.J. Zhu, \NPB{356}{1991}{245}.\nl
J.L. Gervais and Y. Matsuo, \PLB{274}{1992}{309} ({\tt hep-th/9110028});
\CMP{152}{1993}{317} ({\tt hep-th/9201026}).\nl
J.M. Figueroa-O'Farrill, E. Ramos and S. Stanciu,
\PLB{297}{1992}{289},
({\tt hep-th/9209002}).}
\refdef[partcurv]{R.D. Pisarski, \PRD{34}{1986}{670}.\nl
M.S. Plyushchay, \PLB{253}{1991}{50}.\nl
C. Batlle, J. Gomis, J.M. Pons and N. Rom\' an-Roy,
\JPA{21}{1988}{2693}.}
\refdef[Plyus]{M.S. Plyushchay, \MPLA{4}{1989}{837}.}
\refdef[Radul]{A.O. Radul, \JETPL{50}{1989}{371}; \FAaIA{25}{1991}{25}.}
\refdef[GraciaPons]{X. Gr\`acia and J.M. Pons, \JPA{25}{1992}{6357}.}
\refdef[Dickey]{L.A. Dickey, \CMP{87}{1982}{127}.}
\refdef[Zoller]{D. Zoller, \PRL{65}{1990}{2236}.}

%%%%%%%%%%%%%%%%%%%%%%%%

%
%\draftmode
\overfullrule=0pt
\def\pubblock{ \line{\hfil\rm QMW--PH--94-16}
               \line{\hfil\tt hep-th/9406072}
               \line{\hfil\rm June 1994}}
\titlepage
\title{The $\W_3$-particle}
\author{Eduardo Ramos\email{E.Ramos@qmw.ac.uk}}
\andauthor{Jaume Roca\email{J.Roca@qmw.ac.uk}}
\QMW
\abstract{We show that $\W_3$ is the algebra of symmetries of the
``rigid-particle'', whose action is given by the integrated
extrinsic curvature
of its world line. This is easily achived by showing that its equation
of motion can be written in terms of the Boussinesq operator. We also
show how to obtain the equations of motion of the standard
relativistic particle provided it is consistent to impose the
``zero-curvature gauge'', and comment about its connection with the
$\KdV$ operator.}
\endtitlepage
\subsection{Introduction}

The geometrical interpretation of
$\W$-type symmetries has attracted the attention of many
mathematical physicists in recent years. Although a plethora of
interesting results are now at our disposal it is commonly agreed
that we have not yet a complete understanding of the underlying geometry.
It is clear that simple mechanical systems enjoying $\W$ symmetry
could be an unvaluable tool in this difficult task. On one hand they
could provide us with some geometrical and/or physical interpretation
for $\W$-transformations ($\W$-morphisms), while on the other hand
they could give us some hints about which are the relevant structures
associated with $\W$-gravity -- the paradigmatic example being
provided by
the standard relativistic particle and diffeomorphism invariance
($\W_2$).

It is well known by now the connection between $\W$-morphisms and
the extrinsic geometry of curves and surfaces \[extgeo].
Therefore, it seems natural to look
for a $\W$-particle candidate among the geometrical actions depending
on the extrinsic curvature. The canonical analysis of those models
has been carried out by several authors \[partcurv]. In particular
M.S.~Plyushchay studied in \[Plyus] the action given by
$$S=\alpha \int \sqrt{|\kappa^2|} \; ds, \(action)$$
where $\alpha$ is a dimensionless coupling constant\fnote{In our conventions
the coordinates are dimensionless.} and the extrinsic curvature
$\kappa$ is given by
$$\kappa^2 = g_{\mu\nu}{d^2 x^{\mu}\over {d s^2}}
{d^2 x^{\nu}\over {d s^2}},\(curvature)$$
where $g_{\mu\nu}$ stands for minkowskian, euclidean, or any
$x$-independent metric.
He showed that this dynamical system
posseses two gauge invariances, and that one of them is
the expected invariance under diffeomorphisms. It is the main
purpose of this note to show that these gauge invariances are nothing
but $\W_3$.

\subsection{Gauge structure analysis}

We now briefly review the main results concerning the gauge
structure of the action \(action).

In an arbitrary parametrization $x^\mu(t)$ the lagrangian
is given by
$$
L=\alpha\sqrt{\left|{\ddxperp^2}\over{\dot x^2}\right|},\(lagran)
$$
where $\dot x^\mu=dx^\mu/dt$ and $\ddxperp^\mu=\ddot x^\mu
-\dot x^\mu(\ddot x\dot x)/\dot x^2$.

For the study of the system in the hamiltonian framework
it is convenient to introduce the velocity $q^\mu=\dot x^\mu$
as an auxiliary variable in order to avoid the presence of
second-order derivative terms in the lagrangian. The
lagrangian in these new variables reads
$$L =\alpha\sqrt{\left|{\dqperp^2}\over{q^2}\right|} + \gamma
(q - \dot x).\(constlagran)$$

We introduce the canonical momenta $(P_\mu,p_\mu,\pi_\mu)$
associated with the coordinates $(x^\mu,q^\mu,\gamma^\mu)$.
Their Poisson bracket algebra is given by
$$
\{{\cal Q}^\mu,{\cal P}_\nu\}=\delta^\mu_{\;\nu}.\(poissonbra)
$$
The definition of the momenta implies a number of primary hamiltonian
contraints.
Being the lagrangian linear in $\dot x$ we obtain the trivial
the second-class constraints $\gamma^\mu=-P^\mu$ and $\pi^\mu=0$,
from which we can eliminate the canonical pair $(\gamma,\pi)$
by means of the Dirac bracket.
The primary first-class contraints are
$$
\phi_1=pq\approx0,\quad\quad\phi_2={{1}\over{2}}\left(|p^2|-{{\alpha^2}\over
{|q^2|}}\right)\approx0.
\(primfcconst)
$$
Stabilization of these constraints through the standard Dirac
procedure leads to the secondary
$$
\phi_3=Pq\approx0,\quad\quad\phi_4=Pp\approx0,
\(secfcconst)
$$
and tertiary
$$
\phi_5=P^2\approx0,
\(tertfcconst)
$$
first-class hamiltonian constraints, which form a closed
Poisson bracket algebra.

In the constrained submanifold time evolution is generated
by the hamiltonian
$$
H=\phi_3+v_1\phi_1+v_2\phi_2.
\(hamiltonian)
$$
The functions $v_1$ and $v_2$ have a definite expression in terms
of lagrangian quantities,
$$
v_1={{q\dot q}\over{q^2}},\quad\quad
v_2={{1}\over{\alpha}}\sqrt{|q^2\dqperp^2|},
\(vi)
$$
but they should be regarded as arbitrary
functions of time in the canonical formalism. Time evolution
becomes unambiguous once we assign them definite values, which is
nothing but a choice of gauge.
The hamiltonian stabilization procedure guarantees the consistency
of the expressions chosen for the $v_i$ with their lagrangian
expression \(vi).

The presence of two arbitrary
functions reveals the existence of a second gauge symmetry in addition
to the familiar reparametrization invariance.
We shall show in what follows that such symmetries are
precisely $\W_3$.

\subsection{ Equations of motion, Boussinesq operator and
$\W_3$ symmetry}

The hamiltonian equations of motion can be sugestively written as
$$\eqalign{
\dot x^\mu&=q^{\mu},
\cr
\left(\matrix{
\dot P^\mu\cr
\dot p^\mu\cr
\dot q^\mu\cr}\right) & =
\left(\matrix{0&0&0\cr
-1&-v_1&-{\alpha^2 v_2\over {q^4}}\cr
0&v_2&v_1\cr}\right)
\left(\matrix{P^\mu\cr p^\mu\cr q^\mu\cr}\right).
}\(eqmotion)$$
Notice that, neglecting the first equation,
which simply states the definition of $q^{\mu}$, the equations of
motion can be casted in the ``Drinfeld-Sokolov'' form
associated with a particular $SL(3)$ connection. We
will come back to this point later on.

It will be convenient in what follows to introduce a different and
more geometrical parametrization for $v_1$ and $v_2$
Let us define $e$ and $\lambda$ as
$$
e^2=
|q^2|,\quad\quad\quad\lambda^2=|\kappa^2|.\(repara)
$$

Notice that $e^2$ is nothing
but the modulus of the induced metric, and $\lambda$
is basically the extrinsic curvature.
In terms of $e$ and $\lambda$
the gauge degrees of freedom are given by $v_1 = \dot e/e$ and
$v_2 =\lambda e^3/\alpha$.

{}From \(eqmotion) we can obtain a single equation for the velocity
vector $q^\mu$,
$$
\dddot q^\mu+u_1\ddot q^\mu+u_2\dot q^\mu+u_3q^\mu=0,
$$
with
$$\eqalign{
u_1&=-2{{d\ln(e^3 \lambda )}\over{dt}},
\cr
u_2&={e^4\lambda^2}+15{{\dot e^2}\over{e^2}}+7{{\dot e\dot
\lambda}\over{e\lambda}}+2{{\dot\lambda^2}\over{\lambda^2}}
-4{{\ddot e}\over{e}}-{{\ddot\lambda}\over{\lambda}},
\cr
u_3&=e^4\dot\lambda\lambda + e^3\dot e\lambda^2
-15{{\dot e^3}\over{e^3}}-7{{\dot e^2\dot\lambda}\over{e^2\lambda }}
-2{{\dot e\dot\lambda^2}\over{e\lambda^2}}
+10{{\dot e\ddot e}\over{e^2}}+2{{\ddot e\dot\lambda}\over{e\lambda}}
+{{\dot e\ddot\lambda}\over{e\lambda}}-{{\dddot e}\over{e}}.\cr
}\(choris)$$
Notice that $u_1$ is a total derivative. This
allows us to remove the term with the second derivative by a simple
local rescaling of $q^\mu$. Indeed, if we define
$$
y^\mu={1\over e^2\lambda^{2\over 3}}q^\mu,\(redefinition)
$$
the equation for the new velocity vector $y^\mu$ can be written
in terms of the Boussinesq Lax operator
$$
\dddot y^\mu+T\dot y^\mu+(W+{{\dot T}\over2})y^\mu=0,\(Boussequation)
$$
with $T$ and $W$ given by
$$
\eqalign{
T=&\; {e^2\lambda^2}-{1\over3}{\dot e^2\over e^2}-
{\dot e\dot\lambda\over e\lambda}-{4\over3}{\dot\lambda^2\over
\lambda^2}+ 2{\ddot e\over e}+{\ddot\lambda\over\lambda},
\cr
W=&\; e \dot e \lambda^2 + 3 {\dot e^3\over e^3} +{5\over 3}
e^2 \lambda\dot\lambda + {\dot e^2\dot\lambda\over e^2\lambda} +
{4\over 3}{\dot e\dot\lambda^2\over e \lambda^2} +{56\over 27}
{\dot\lambda^3\over\lambda^3}\cr
&-4{\dot e \ddot e\over e^2} -{2\over 3}{\ddot e\dot\lambda\over e
\lambda}-{\dot e\ddot\lambda\over e\lambda} -{8\over 3}
{\dot\lambda\ddot\lambda\over \lambda^2} + {\dddot e\over e}+
{2\over 3}{\dddot\lambda\over\lambda}.
}$$

The algebra of symmetries of equations of the type $\L \Psi =0$ with
$\L$ a differential operator of the form $\partial ^n + ...$ has been
studied by Radul in \[Radul]. From his general construction it is easily
deduced that for the particular case of $\L = \d^3 + T\d + W + \dot
T/2$, which is our case of interest, the symmetry algebra is $\W_3$.
Rather than reproducing here the general arguments leading to this
result, we will work out explicitly, for convenience of the
less mathematically oriented reader, the case at hand.

The most general (local) variation of $\Psi$ preserving the
structure of $\L$ , up to
equations of motion, is given by
$$\delta_{\epsilon ,\rho} =
\rho \ddot\Psi + (\epsilon -{1\over 2}\dot\rho )\dot\Psi
-(\dot\epsilon +{1\over 6}\ddot\rho + {2\over 3} \rho T)\Psi,
\(varpsi)$$
where the parametrization has been chosen so that $\epsilon$ and
$\rho$ denote the parameters associated with diffeomorphisms and
pure $\W_3$-morphisms respectively. The corresponding transformations
for $T$ and $W$ which leave the equations of motion invariant are
given by
$$\eqalign{
\delta^{(T)}_{\epsilon}T = & \; 2 \dddot\epsilon + 2 \dot\epsilon T
+\epsilon\dot T,\cr
\delta^{(T)}_{\epsilon}W = & \; 3\dot\epsilon W +\epsilon\dot W,\cr
\delta^{(W)}_{\rho} T = &\; 3 \dot\rho  W + 2\rho \dot W,\cr
\delta^{(W)}_{\rho} W = & \; -{1\over 6} \rho^{({\rm v})}
- {5\over 6} \dddot\rho
T -{5\over 4}\ddot\rho\dot T - \dot\rho ({3\over 4}\ddot T
+{2\over 3}T^2) -\rho ({1\over 6} \dddot T +{2\over 3} T\dot T).\cr
}\(wmorphisms)$$

It is now a long but straightforward exercise to check that these
transformations obey the algebra of $\W_3$ transformations, {\ie}
$$\eqalign{
\comm{\delta^{(T)}_{\epsilon_1}}{\delta^{(T)}_{\epsilon_2}} = & \;
\delta^{(T)}_\epsilon,\quad{\rm where}\quad
\epsilon=\epsilon_1\dot\epsilon_2 -\dot\epsilon_1\epsilon_2,
\cr
\comm{\delta^{(T)}_{\epsilon}}{\delta^{(W)}_{\rho}} = & \;
\delta^{(W)}_{\tilde\rho},\quad{\rm where}\quad
\tilde\rho=\epsilon\dot\rho - 2\dot\epsilon\rho,
\cr
\comm{\delta^{(W)}_{\rho_1}}{\delta^{(W)}_{\rho_2}} = & \;
\delta^{(T)}_\epsilon,\quad{\rm where}\quad
\epsilon={{2\over 3}\rho_1\dot\rho_2 T -{1\over 4}\dot\rho_1
\ddot\rho_2 +{1\over 6}\rho_1\dddot\rho_2 - (\rho_1\lra\rho_2)},\cr
}\(commutators)$$
where we have assumed for simplicity that the parameters are field
independent. The most general case can be equally worked out from
the results of \[Radul].

Notice that the consistency of the procedure in our case
is based on the fact that
$T$ and $W$, through their dependence in $\lambda$ and $e$, are gauge
degrees of freedom themselves, therefore before any solution to the
equations is found they should be given definite values. In this
language the $\W$-morphisms given by \(wmorphisms) are to be
understood as gauge transformations.

Because the relationship between $y^\mu$ and $q^{\mu}$ is not
easily invertible the previous analysis does not yield directly
the transformation rules for $q^\mu$. These transformations
are interesting by themselves because the action is naturally
written in terms of $q^\mu$. Here is where standard techniques
in constrained dynamical systems come to our rescue.
The gauge transformations for $x^\mu$, $q^\mu$, and $\gamma^\mu$
can be computed
using the general method of \[GraciaPons] to give
$$\eqalign{
\delta_{\beta,\eta}x^\mu &=\beta q^\mu +
\dot\eta {\alpha\over e^3 \lambda}\dot q^\mu_{\perp}- 2\eta\gamma^\mu,\cr
\delta_{\beta,\eta}q^\mu &= \dot\beta q^\mu + \beta\dot q^\mu
-\dot\eta\left( \gamma^\mu +\alpha {\lambda\over e} q^\mu +
\alpha {\dot e\over e^4\lambda}\dot q^\mu_{\perp}\right) +
\ddot\eta {\alpha \over e^3\lambda}\dot q^\mu_{\perp},\cr
\delta_{\beta,\eta}\gamma^\mu &=0.\cr
}\(varq)$$
A direct computation now shows
that the tranformations given by \(varq) are not
only a symmetry of the action, but also that, as a soft algebra,
reproduce the algebra of $\W_3$ transformations if we do the following
identifications:
$$\eqalign{
\epsilon &=\beta + {\alpha\over 2}{\ddot\eta\over e^3 \lambda}
+\dot\eta {\alpha\over e^3\lambda}\left( {1\over 6}{\dot\lambda
\over \lambda} - {1\over 2}{\dot e\over e}\right),\cr
\rho &= - {\alpha\over e^3\lambda}\dot\eta.\cr}\(changepara)$$
Notice that in this parametrization the variation of $x^\mu$ has a
nonlocal expression. This is not, to some extent, surprising due to
the
fact that the natural variable for the system seem to be supplied by
$q^\mu$ rather than the coordinates themselves\fnote{This is
somehow reminiscent of what happens for the case of the two
dimensional massless boson, where the natural variable is supplied
by its associated $U(1)$ current.}.
Moreover, as expected,
the variations induced on $y^\mu$ through \(varq)
coincide with the ones obtained via \(varpsi), tightening up the
formalism, and showing the nice interplay between the hamiltonian
and the $\W$-algebraic methods of tackling the problem.

There is also a more indirect, but interesting in its own, way to
show the invariance of the action \(action) under $\W_3$-morphisms
with the help of the Miura transformation. In order to do so we
should return to the expression of the hamiltonian equations of
motion in terms of the $SL(3)$ connection. If we write \(eqmotion) as
$$\left( {{d\ }\over {dt}} - \Lambda \right)\left(\matrix{P^\mu\cr
p^\mu\cr q^\mu\cr}\right)=0,\(DrinSok)$$
it is clear that this equations have local gauge invariance
under
$$\left(\matrix{P^\mu\cr p^\mu\cr q^\mu\cr}\right)\to
M \left(\matrix{P^\mu\cr p^\mu\cr q^\mu\cr}\right)\quad
{\rm and}\quad \Lambda\to M\Lambda M^{-1} - M { dM^{-1}\over dt },$$
with $M\in SL(3)$.
It will be convenient for our purposes to work in the Miura gauge,
{\ie}
$$\Lambda = \left(\matrix{-\phi_1 -\phi_2&0&0\cr
1&\phi_2&0\cr 0&1&\phi_1}\right).\(miuragauge)$$
The matrix $M\in SL(3)$ which brings $\Lambda$ to this form is
$$M ={\lambda^{1\over 3}\over\alpha^{1\over 3}}
\left(\matrix{e&0&0\cr
0&-e&-i {\alpha}/e\cr
0&0& -\alpha/e^2\lambda}\right).$$
Some straightforward algebra now yields
$$\eqalign{\phi_1 &= -{\dot e\over e} -{2\over 3}{\dot\lambda
\over\lambda} - i {\lambda e}\cr
\phi_2 &= {1\over 3}{\dot\lambda \over\lambda}
+ i {\lambda e}\cr}.\(miurafields)$$
The key point, for our present interest, is given by the fact that
the lagrangian \(lagran) can be written as
$$L \sim {\rm Im}(\phi_1 ) \sim {\rm Im}(\phi_2 ).\(lagmiura)$$
The transformation under $\W_3$-morphisms of the Miura fields can
be computed using the Kupershmidt-Wilson
theorem \[Dickey]. If we denote by $\epsilon$
the parameter associated with diffeomorphisms, and by $\rho$ the
one associated with pure $\W_3$-morphisms, these variation
are given by
$$\eqalign{\delta_{\epsilon ,\rho }\phi_1 =
{{d\ }\over dt} & \left( \dot \epsilon + \epsilon\phi_1 +
{1\over 6} \ddot\rho +{1\over 2}(\dot\phi_1\rho
+\phi_1\dot\rho)\right.
\cr
& \left.-{1\over 3}({1\over 2}\dot\phi_1 -2\dot\phi_2 -\phi_1^2 + 2
\phi_2^2 + 2\phi_1\phi_2 )\rho\right),\cr
\delta_{\epsilon ,\rho}\phi_2 = {{d\ }\over dt} &
\left( \epsilon\phi_2 -
{1\over 3} \ddot\rho -{1\over 2}(\dot\phi_2\rho +\phi_2\dot\rho
) -\dot\phi_1\rho +\phi_1\dot\rho\right.
\cr & \left.
+{1\over 3}(\dot\phi_1 +{1\over 2}\dot\phi_2 - 2\phi_1^2 +
\phi_2^2 - 2\phi_1\phi_2 )\rho\right),\cr
}\(transfos)$$
which is a total derivative! And from this follows directly the
invariance of the action.

\subsection{The relativistic particle}

It is easy to recover the equations of motion for the standard
relativistic particle from \(eqmotion) whenever it is consistent to
impose the gauge condition $v_2 =0$. Notice that since $v_2$
is proportional to the extrinsic curvature, this
gauge can only be consistently imposed when it can be taken to be zero
(initial conditions may be incompatible with this gauge choice).
In that case the
hamiltonian equations of motion collapse to
$$ \ddot x^\mu - {\dot e\over e}\dot x^\mu =0,\(relpartequ)$$
which is the equation of motion for the relativistic particle.
Notice that \(relpartequ) can be sugestively written
as $\ddot x^\mu_{\perp} = 0$.

It is natural to ask now, in the light of our previous discussion,
which role, if any, is played by the $\KdV$ Lax operator in this
case. The answer to this question is easily obtained by realising
that via the redefinition
$$ x^\mu\to e^{{1\over 2}}x^\mu\(redefpart)$$
the equations of motion for the new variable can be written as
$$\ddot x^\mu + T x^\mu =0,\(kdvform)$$
with
$$T={1\over 2}{\ddot e\over e} -{3\over 4}{\dot e^2\over e^2},
\(Tpar)$$
and being $e$ a gauge degree of freedom for the relativistic
particle we can apply all the same arguments as above.

\subsection{Final comments and digressions}

We hope to have convinced the reader that $\W_3$-symmetry does play
an important role in the symmetry structure of the
rigid particle. And that standard techniques in constrained hamiltonian
systems and $\W$-algebras can be intertwined as powerful tools for
the better understanding of both disciplines. But, of course,
much is still
to be done. For example, we do not yet completely understand the
appearance of the $SL(3)$ structure in the problem, which should be
the key to generalize these procedures to other $\W$-algebras.
Moreover, under quantization \[Plyus]
the rigid particle is associated
with massless representations of the Poincare group
with integer helicity, therefore being a potential candidate for
a particle description of photons, gravitons and higher spin fields.
It is a tantalizing possibility that $\W_3$ symmetry can play a role
in such physical systems.

We would not like to finish without a few sentences about the possible
relevance of our results to $\W_3$-gravity. It was shown in \[Zoller]
that a naive coupling to gravity of the action \(action), {\ie}
to consider arbitrary $x$-dependent metrics, yields
inconsistencies. This is due to the fact that the constraint algebra
no longer closes in curved space-time. We believe this to be a signal
indicating that the $\W_3$-particle is only consistently coupled to
$\W_3$-gravity. Of course this obscure sentence needs some explanation.
In the standard particle case the relevant bundle is the tangent
bundle of the manifold, $TM$. In this case it is well known how to equip
the bundle with a metric structure and all the powerful and well
understood machinery of Riemmanian geometry is at our disposal.
In the language of jet bundles $TM$ is nothing but $J^1(\reals , M)$, and
it is our understanding that the relevant bundle for the $\W_3$-particle
is provided by $J^2(\reals ,M)$, which is not itself a vector bundle.
It is well known that $J^2(\reals , M)$ is an associated bundle to
$F^2M $, the frame bundle of second order, and it is our belief that
the required structure should be a ``natural'' structure in $F^2 M$,
as it is the metric in $FM$. It is our hope that the action of the
$\W_3$-particle will provide us with a ``$\W_3$-line element'', thus
offering some valuable insight about which generalized
structures in $F^2M$ should be considered.
Work on this is in progress.

\ack
We would like to thank J.M. Figueroa-O'Farrill for many useful
discussions about the $\W$-orld. One of us, J.R. (no pun intended),
is also thankful to the spanish ministry of
education for financial support.

\refsout
\bye